\documentclass[conference]{IEEEtran}
\IEEEoverridecommandlockouts
\usepackage{cite}
\usepackage{amsmath,amssymb,amsfonts}
\usepackage{algorithmic}
\usepackage{graphicx}
\usepackage{textcomp}
\usepackage{censor}
\usepackage{multirow}
\usepackage{makecell}
\usepackage{arydshln}
\usepackage{listings}
\usepackage{soul}
\usepackage{censor}

\usepackage{array}
\usepackage{ragged2e}
\usepackage{enumitem}
\usepackage{url}
\usepackage{wrapfig}
\usepackage{balance}
\usepackage{colortbl}
\usepackage{soul,color}
\usepackage{multirow}
\usepackage[table,xcdraw]{xcolor}
\usepackage[most]{tcolorbox}

\usepackage{hyperref}
\usepackage{flushend}

\hypersetup{
    colorlinks=true,
    linkcolor=blue,
    filecolor=blue,      
    urlcolor=blue,
    citecolor=blue
    }

\newtcolorbox{myquote}[1][]{%
    colback=black!5,
    colframe=black!5,
    notitle,
    sharp corners,
    borderline west={2pt}{0pt}{gray!80!black},
    enhanced,
    breakable,
    }

\definecolor{orange}{rgb}{1,0.647,0}

\usepackage[utf8]{inputenc}
\usepackage{fourier}

\makeatletter
\newcommand\footnoteref[1]{\protected@xdef\@thefnmark{\ref{#1}}\@footnotemark}
\makeatother

\usepackage[most]{tcolorbox}
\usetikzlibrary{calc}

\newcolumntype{L}[1]{>{\raggedright\let\newline\\\arraybackslash\hspace{0pt}}m{#1}}
\newcolumntype{C}[1]{>{\centering\let\newline\\\arraybackslash\hspace{0pt}}m{#1}}
\newcolumntype{R}[1]{>{\raggedleft\let\newline\\\arraybackslash\hspace{0pt}}m{#1}}

\newcommand{\fig}[1]{Fig.~\ref{fig:#1}}

\usepackage{lipsum}                     
\usepackage{xargs}                      
\usepackage[colorinlistoftodos,prependcaption,textsize=scriptsize]{todonotes}

\usepackage{multirow}
\usepackage{booktabs}
\usepackage{bigdelim}
\usepackage{tikz}
\usepackage{enumitem}

\def\BibTeX{{\rm B\kern-.05em{\sc i\kern-.025em b}\kern-.08em
    T\kern-.1667em\lower.7ex\hbox{E}\kern-.125emX}}

\begin{document}

\title{Comparing 2D and Augmented Reality Visualizations for Microservice System Understandability: A Controlled Experiment
}



\author{\IEEEauthorblockN{
Amr S. Abdelfattah}
\IEEEauthorblockA{\textit{Computer Science of Baylor University} \\
Waco, TX, USA \\
amr\_elsayed1@baylor.edu}
\and
\IEEEauthorblockN{\hspace{.5em}}
\and
\IEEEauthorblockN{
Tomas Cerny}
\IEEEauthorblockA{\textit{Computer Science of Baylor University} \\
Waco, TX, USA \\
tomas\_cerny@baylor.edu}
\and
\IEEEauthorblockN{\hspace{.5em}}
\and
\IEEEauthorblockN{
Davide Taibi}
\IEEEauthorblockA{\textit{Tampere University} 
Tampere, Finland}
\IEEEauthorblockA{\textit{University of Oulu} 
Oulu, Finland \\
davide.taibi@oulu.fi}
\and
\IEEEauthorblockN{\hspace{3em}}
\and
\IEEEauthorblockN{
Sira Vegas}
\IEEEauthorblockA{\textit{Universidad Politecnica de Madrid} \\
Madrid, Spain \\
svegas@fi.upm.es}
}


\maketitle

\begin{abstract}
Microservice-based systems are often complex to understand, especially when their sizes grow. Abstracted views help practitioners with the system understanding from a certain perspective. Recent advancement in interactive data visualization begs the question of whether established software engineering models to visualize system design remain the most suited approach for the service-oriented design of microservices.
Our recent work proposed presenting a 3D visualization for microservices in augmented reality. This paper analyzes whether such an approach brings any benefits to practitioners when dealing with selected architectural questions related to system design quality. For this purpose, we conducted a controlled experiment involving 20 participants investigating their performance in identifying service dependency, service cardinality, and bottlenecks. Results show that the 3D enables novices to perform as well as experts in the detection of service dependencies, especially in large systems, while no differences are reported for the identification of service cardinality and bottlenecks. We recommend industry and researchers to further investigate AR for microservice architectural analysis, especially to ease the onboarding of new developers in microservice~projects.
\end{abstract}


\section{Introduction}
Cloud-native design is associated with complex systems involving many self-contained microservices that interact to solve enterprise problems. One of the great benefits is decentralization, which enables independent teams to develop and maintain self-contained microservices providing better autonomy and independent deployment. At the same time, it is connected with one major challenge for microservices; with decentralization, we lose the system-centric perspective that aids observability and holistic system understanding. Such a perspective would be very useful to help drive efficient system evolution.

Various architectural views for the system-centric perspective can greatly abstract the underlying system complexity. They can reduce unnecessary details when reasoning about microservice dependencies or other perspectives (i.e., domain, technology, operation). We can analyze existing systems and extract these views to better analyze their architecture across their evolution~\cite{10.1007/978-3-030-49418-6_21}. This extraction can be performed manually or involve dynamic or static analysis. Given that microservices emerged from service-oriented architecture, the system's service view is well adopted by many existing tools used for monitoring cloud-native systems, and thus we focus on this view. 

However, existing tools use an all-in-one static service view visualization lacking interactivity~\cite{cerny2022microservice, 9944666}. Moreover, we must assume that the system size in terms of different microservices can grow beyond what established visualization models had been designed for (i.e., monolithic systems). 
The question is whether Augment Reality (AR) and its interactivity shift could make practitioners perform better than when using the established visualization when solving common quality reasoning tasks in service views applied to microservices.


In this paper, we elaborate on the service view visualization. Our recently proposed service view visualization approach uses a 3D node-edge model rendered in Augmented Reality (AR)~\cite{cerny2022microvision}. 
This visual model has been proposed with the rationale that 3D space can better accommodate complex graphs, and it might be easier for end users to read it within limited space rather than using the established conventional 2D visual models. 

The goal of our work is to compare the perceived understandability of microservice-based systems while analyzing the service call graph visualized in 2D or AR. 
For this purpose, we conduct a controlled experiment among 20 developers comparing how they perform when detecting service dependencies, service cardinality, and bottlenecks in two different sizes variations of a system.

Moreover, we also aim at understanding  how practitioners would accept the new AR model and what challenges emerged from using the AR to better understand the impact of the technological~shift.

The results conclude that AR visualization enables novice developers to understand system dependencies as if they were experienced developers. Also, it shows a large impact for visualizing large systems compared with the 2D tool, which begins to degrade accordingly.


The results of this work can be useful to researchers that could further develop and evaluate the AR visualizations for microservices, but also for industry that can extend existing tools to integrate AR into their tools. As an example, tracing tools, such as Jaeger (\url{https://www.jaegertracing.io}) or Kiali (\url{https://kiali.io}), might easily integrate AR on top of their 2D service view.    

This paper is organized as follows. Section II details related work and microservice visualization. Section III states the research questions and study variables used to answer them. Section IV details the study with its experimental design details. Study results are presented in Section V, and discussed in Section VI. Threats to the validity of this study are given in Section VII and conclusions in Section VIII.


\section{Related Works}

The process of Software Architecture Reconstruction (SAR)~\cite{o2002software} aims to analyze an existing system to obtain the implemented architecture of the system. The architecture is a system blueprint and is central to the development and design of component-based systems~\cite{o2002software}.
According to O'Brien et al.~\cite{o2002software}, the outcome of the process can be used to evaluate the conformance of the as-built architecture to the as-documented architecture; for systems that are poorly documented; or analyzing and understanding the architecture of existing systems to enable modification of the architecture to satisfy new requirements and to eliminate existing software deficiencies, which can be seen a system's evolvability~\cite{bogner2021}. In addition, this process has been used on systems for verification, and trade-off analysis~\cite{10.1007/978-3-030-49418-6_21}.

For SAR in microservices~\cite{8417116, 10.1007/978-3-030-49418-6_21,recovering_architecture}, we find relevant system information in the system codebases, source code, even container descriptors, build files, or many configuration files. In manual reconstructions, even existing documentation can be used~\cite{8417116, 10.1007/978-3-030-49418-6_21}.
We can also use system runtime and monitoring to determine dependencies across microservices~\cite{Thalheim:2017:SAI:3135974.3135977,Esparrachiari:2018:TCM:3277539.3277541}. Microservice dependency graphs have been used to test microservices~\cite{service_dependency}, which shows that dynamic analysis could provide a viable path to analyze systems holistically from certain perspectives.

Software architecture might mean different things to different experts, which is why we typically consider different viewpoints. Several such views have been introduced for the microservice architecture~\cite{8417116, 10.1007/978-3-030-49418-6_21} such as \textit{Domain View} that deals with concerns of domain experts through domain models and microservice bounded contexts; \textit{Technology View} that identify applied technologies in microservices; \textit{Service View} that constructs service models that specify microservices, interfaces, and endpoints; and \textit{Operation View} that concerns with the topology for service deployment and the infrastructure for, service orchestration, discovery, monitoring, and other operations concerns.

While there are broad options to perform SAR, we do not always need a comprehensive architectural overview. It might be sufficient to focus on a selected viewpoint. For instance, it might be sufficient to analyze microservice dependency graphs to test them~\cite{service_dependency}, and for this, we may use a simplified process. Similarly, existing tracing and monitoring tools typically provide a primitive service overview (i.e., service call graph). It is a natural next step to visualize the overall system to practitioners in a tailored way. Therefore, they can better understand the system for the tasks they use these tools for. Thus, the derived experiment measures the impact of different visualization methods on the system's understandability for practitioners.


\subsection{Microservice Visualization}


The emerging question in the context of microservices is how to visualize the system viewpoints. Conventional visual models like UML and SysML are derivatives, but microservices add various aspects to consider, primarily system size. The question is whether the existing visualization techniques used before the microservice architecture are appropriate for the present challenges~\cite{10015027,9944666}. Although we can effortlessly apply UML class diagrams to represent the domain in the context of microservice, it is still a matter of concern how to portray service views. The Open Group Architecture Framework (TOGAF)~\cite{open2002open} and ArchiMate enterprise architecture modeling language~\cite{lankhorst2010anatomy} predate microservices by almost two decades.

Microservice researchers observed that when it comes to microservices, the system holistic view should include service APIs and their interactions~\cite{mayer_weinreich_dashboard}. The SAR's \textit{service view} describes the APIs of microservices and the inter-service calls between them.
Furthermore, the service view might help both DevOps and developers understand dependencies inter-connections to assess possible ripple effects in their evolution~\cite{bogner2021}. 

Data visualization has introduced alternative avenues which could be utilized; these include 3D node-edge graphs used in semantic webs~\cite{redgraph}, virtual reality with various metaphors  (i.e., islands, cities) in systems and distributed systems~\cite{vr_distributed,vrea},
and augmented reality~\cite{ar_software_islands}.
Microservice architecture, specifically the service view and the emerging system sizes, seems to lead toward more efficient space utilization, which can be provided by these approaches.

One such approach applied to microservices is Microvision~\cite{cerny2022microvision}, which combines 3D node-edge graphs with AR. 
Microvision considers  the  endpoints of each service and connects services that interact over these endpoints. These can be obtained by tracing tools as well as static analysis tools like Open API (\url{https://www.openapis.org}, formally Swagger 
 \url{https://swagger.io}) when combined with the identification of remote method calls through REST templates or similar. 
Dynamic analysis could be more precise in identifying the actual remote calls. This is because the static analysis approach has limits when considering matching call and endpoint signatures and HTTP method, possibly augmented with configuration information on target service names). However, dynamic analysis requires comprehensive traffic to identify all trace combinations. 
With incomplete traces, we might not uncover the entire service call graph. The microservice perspective in the service view gives a service call graph representing the system broken down into microservices and specific endpoints, showing how the microservices communicate with one another. 

Microvision operates in AR space, which changes the navigation through the service view from what we are used to with established visual models. Instead of using a personal computer, keyboard, and mouse, we need to use camera-equipped devices to navigate around the rendered system graph. In addition, more interaction is added to such a representation where we can select a given microservice to highlight its connections and hierarchically detail microservice endpoint and their properties. 

The challenge is that one cannot answer whether such an approach can help practitioners to better or more effectively cope with common tasks when identifying architectural properties, qualities, and service dependencies. Moreover, there are likely to arise new practical obstacles this technological shift may introduce. 

\section{Research Questions and Study Variables} \label{sec:rqs}



This study evaluates two microservice-based system visualization approaches: the 2D visualization and the 3D Augmented Reality (AR)-based visualization. Therefore, it examines the impact of the AR approach on the understandability of microservice-based systems. Moreover, the study considers system size and practitioner experience level as the main factors for interaction with those tools. The \textit{system size} shows the influence of the number of microservices on the system understandability. Therefore, the study examines the tools with small and large system variants. On the other hand, the \textit{practitioner experience level} measures the impact of the different development experience levels on the system understandability. The study divides the participants into two groups: novice and experienced, based on their experience level in microservices development. In summary, the goal of the experiment is formulated~as~follows:\newline

\vspace{-2em}
\begin{myquote}
\noindent
\underline{Evaluating} two visualization approaches,
\underline{for the} \underline{purpose of} measuring the microservice-based system \textit{understandability},
\underline{with respect to} the system size and experience level of practitioners.
\end{myquote}

\vspace{-.5em}
This paper composites and addresses the following Research Questions (RQs) to achieve the goal:

\subsection*{\textbf{RQ1. Is 2D visualization more applicable than AR for understanding microservice-based systems?}}

The study is designed to introduce specific tasks that reason about the understandability of different aspects of systems. Therefore, the study evaluates the system understandability through the visualization tools using the following perspectives:

\begin{itemize}
    \item \textbf{Dependency:} It measures the ability to  identify the dependencies between microservices in a system using given visualization tools.
    \item \textbf{Cardinality:} It measures the ability to recognize the degree of dependency between microservices.
    \item \textbf{Bottleneck:} It examines the degree to which the participant is able to detect the most dependant microservices among the whole system.
\end{itemize}
\vspace{-.5em}


\subsection*{\textbf{RQ2. What is the verbalized perception of the participants regarding the use of AR for understanding microservice-based systems?}}

This question pertains to the feedback on tool usage while performing the tasks.

The evaluation process highlights the following criteria for answering this question:



\begin{itemize}
    \item \textbf{Easiness:} It measures the effort needed by the participant to use the tools and how easy to find the answers to the required tasks.
    \item \textbf{Completeness:} It measures the ability to find all the information needed to understand the system and answer the tasks.
    \item \textbf{Recommendation:} It measures the degree to which the participant recommends the tool for the daily work environment.
\end{itemize}
\vspace{-.5em}


\subsection*{\textbf{RQ3. What are the challenges perceived by the participants with regard to AR-based techniques?}}

This question corresponds to the participants' thoughts on the studied tools. Evaluating challenges regarding the study requires the analysis of various measurement criteria. Therefore, the study collects textual opinions on multiple criteria that help in identifying the challenges and recommendations that could be used to overcome them. These criteria are summarized by measuring the painful points in the tools compared with the easiest ones; that is regarding to how much effort is consumed by the participant to bypass the task.

The following sections discuss the study details and explain the answers to these RQs.

\begin{table}[h]
\centering
\vspace{0em}
\caption{Participants Distribution}
\label{tab:participants}
\begin{tabular}{|p{0.14\linewidth}|p{0.15\linewidth}|p{0.1\linewidth}|p{0.1\linewidth}|p{0.1\linewidth}|p{0.1\linewidth}|}
\hline
\textbf{Participant} & \textbf{Experience} & \textbf{System Size} & \textbf{Session 1} & \textbf{Session 2} & \textbf{Group} \\ \hline
P1 & Novice & Large & 2D & \cellcolor{red!25}-- & G1 \\ \hline
P2 & Experienced & Large & 2D & -- & G1 \\ \hline
P3 & Novice & Large & 2D & AR & G1 \\ \hline
P4 & Novice & Large & 2D & AR & G1 \\ \hline
P5 & Novice & Large & 2D & AR & G1 \\ \hline
P6 & Experienced & Small & 2D & -- & G1 \\ \hline
P7 & Experienced & Small & 2D & -- & G1 \\ \hline
P8 & Experienced & Small & 2D & -- & G1 \\ \hline
P9 & Experienced & Large & 2D & AR & G1 \\ \hline
P10 & Experienced & Small & 2D & AR & G1 \\ \hline
\textbf{P11} & Novice & Small & 2D & AR & G1 \\ \hline
P12 & Experienced & Large & 2D & AR & G1 \\ \hline
\textbf{P13} & Experienced & Large & AR & 2D & G2 \\ \hline
P14 & Novice & Large & AR & 2D & G2 \\ \hline
P15 & Experienced & Small & AR & 2D & G2 \\ \hline
\textbf{P16} & Novice & Small & 2D & AR & G1 \\ \hline
P17 & Experienced & Small & AR & 2D & G2 \\ \hline
P18 & Novice & Small & -- & 2D & G2 \\ \hline
P19 & Novice & Large & AR & 2D & G2 \\ \hline
P20 & Experienced & Small & AR & 2D & G2 \\ \hline
\end{tabular}
 \vspace{-2em}
\end{table}

\begin{table*}[h]
\vspace{0pt}
\caption{Study tasks for the full-size system (41 services)}
\label{tab:tasks}
\begin{tabular}{|p{0.5\linewidth}|p{0.5\linewidth}|}
\hline
\textbf{2D Tasks} & \textbf{AR Tasks} \\ \hline
\multicolumn{2}{|l|}{\textbf{Dependency}} \\ \hline
1) List services that depend on the service "ts-order-service." \newline
    \textbf{Answers:} ts-cancel-service, ts-admin-order-service, ts-seat-service, ts-preserve-service, ts-execute-service, ts-inside-payment-service, ts-travel-service, ts-security-service. & 
1) List services that depend on the service "ts-travel-service" \newline \textbf{Answers:} ts-admin-travel-service, ts-route-plan-service, ts-preserve-service, ts-travel-plan-service, ts-seat-service, ts-food-service.
 \\ \hline
2) List services that the service "ts-travel-service" depends on. \newline \textbf{Answers:} ts-ticketinfo-service, ts-seat-service, ts-train-service, ts-route-service, ts-order-service. &
2) List services that the service "ts-order-service" depends on. \newline \textbf{Answers:} ts-station-service.
 \\ \hline
\multicolumn{2}{|l|}{\textbf{Cardinality}} \\ \hline
3) How many services depend on the service "ts-order-service"? \newline \textbf{Answers:} 8 Services. & 
3) How many services depend on the service "ts-travel-service"? \newline \textbf{Answers:} 6 Services.
\\ \hline 
4) How many services are there that the service "ts-travel-service" depends on? \newline \textbf{Answers:} 5 Services. &
4) How many services there are that the service "ts-order-service" depends on? \newline \textbf{Answers:} 1 Service.
\\ \hline
5) Mention two microservices that have at least 4 dependent microservices (depend on it). \newline \textbf{Answers:} ts-seat-service, ts-travel-service, ts-user-service, ts-route-service,
ts-station-service, ts-order-service. &
5) Mention one microservice that has at least 5 dependent microservices (depend on it) and state the exact number of them. \newline \textbf{Answers:} ts-travel2-service, ts-admin-basic-info-service, ts-route-service, ts-preserve-service. \\ \hline
\multicolumn{2}{|l|}{\textbf{Bottleneck}} \\ \hline
6) What is the most dependent microservice? \newline \textbf{Answers:} ts-preserve-service (12 dependencies). &
6) What is the most dependent microservice? \newline \textbf{Answers:} ts-preserve-service (12 dependencies).
\\ \hline
\end{tabular}
\vspace{-1.5em}
\end{table*}


\begin{table}[h]
\caption{Aspects Feedback 5-Point Likert Scale Ranking \\(1 is the worst, 5 is the best)}
\label{tab:feedback-options}
\begin{tabular}{|p{0.96\linewidth}|}
\hline
\textbf{Easiness} \\ \hline
1) Usability (Easy to use) \\ \hline
2) Understandability (Ease to Understand the system) \\ \hline
3) Time to get information \\ \hline
\textbf{Completeness} \\ \hline
4) Has all information needed \\ \hline
\textbf{Recommendation} \\ \hline
5) How likely are you to recommend this method for daily work? \\ \hline
\end{tabular}
 \vspace{-1em}
\end{table}

 

\begin{table}[h]
 \caption{Feedback Paragraph Questions}
 \label{tab:feedback-paragraphs}
\begin{tabular}{|p{0.96\linewidth}|}
\hline
\textbf{Express these questions (Textual Response)} \\ \hline
1) What is the information that you can easily get using this method? \\ \hline
2) What information do you find hard to get from this method? \\ \hline
3) What has been your biggest pain point? \\ \hline
4) What are your recommendations for enhancing this method? \\ \hline
\end{tabular}
\vspace{-1em}
\end{table}

 

\section{Experimental Design}

This study measures two visualization approaches' impact on the understandability of microservice-based systems. The following subsections go through the setup process until reaching the data analysis conclusion. The~study protocol\footnote{\label{foot-protocol}Study Protocol: \url{https://zenodo.org/record/7693694\#.ZAEi\_S2B3RY}.} including materials and scripts were prepared by the authors and reviewed by Baylor Univerisity's Institutional Review Boards (IRBs) (\#1845572).

\begin{figure}[hb]
\vspace{-2em}
\centering
\includegraphics[width=0.49\textwidth]{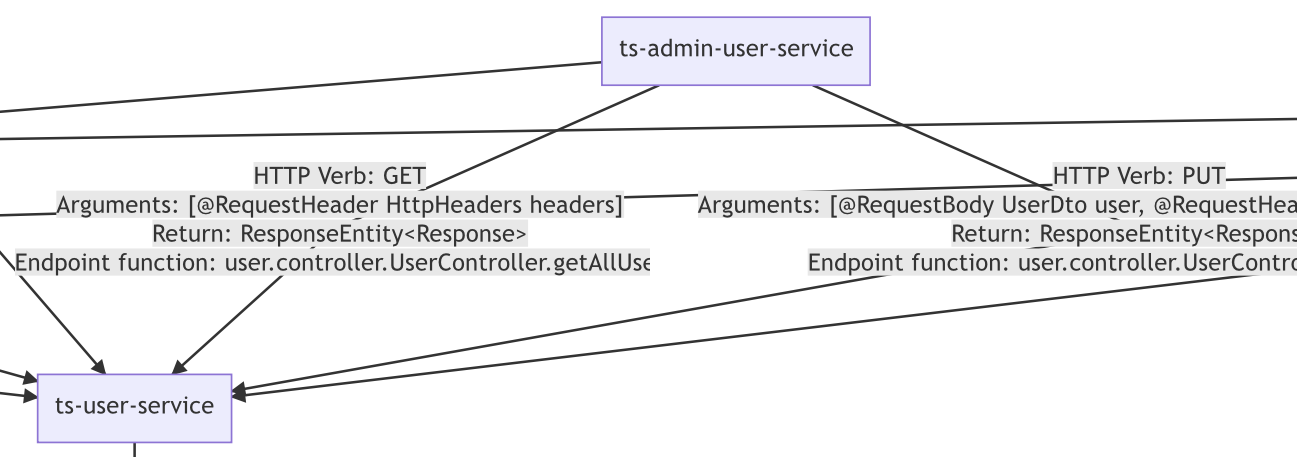}
\vspace{-1.5em}
\caption{2D Tool (Complete version is available at the protocol data.)}
\label{fig:video_sample_2d}
\vspace{-1.5em}
\end{figure}

\begin{figure}[hb]
\centering
\includegraphics[width=0.35\textwidth]{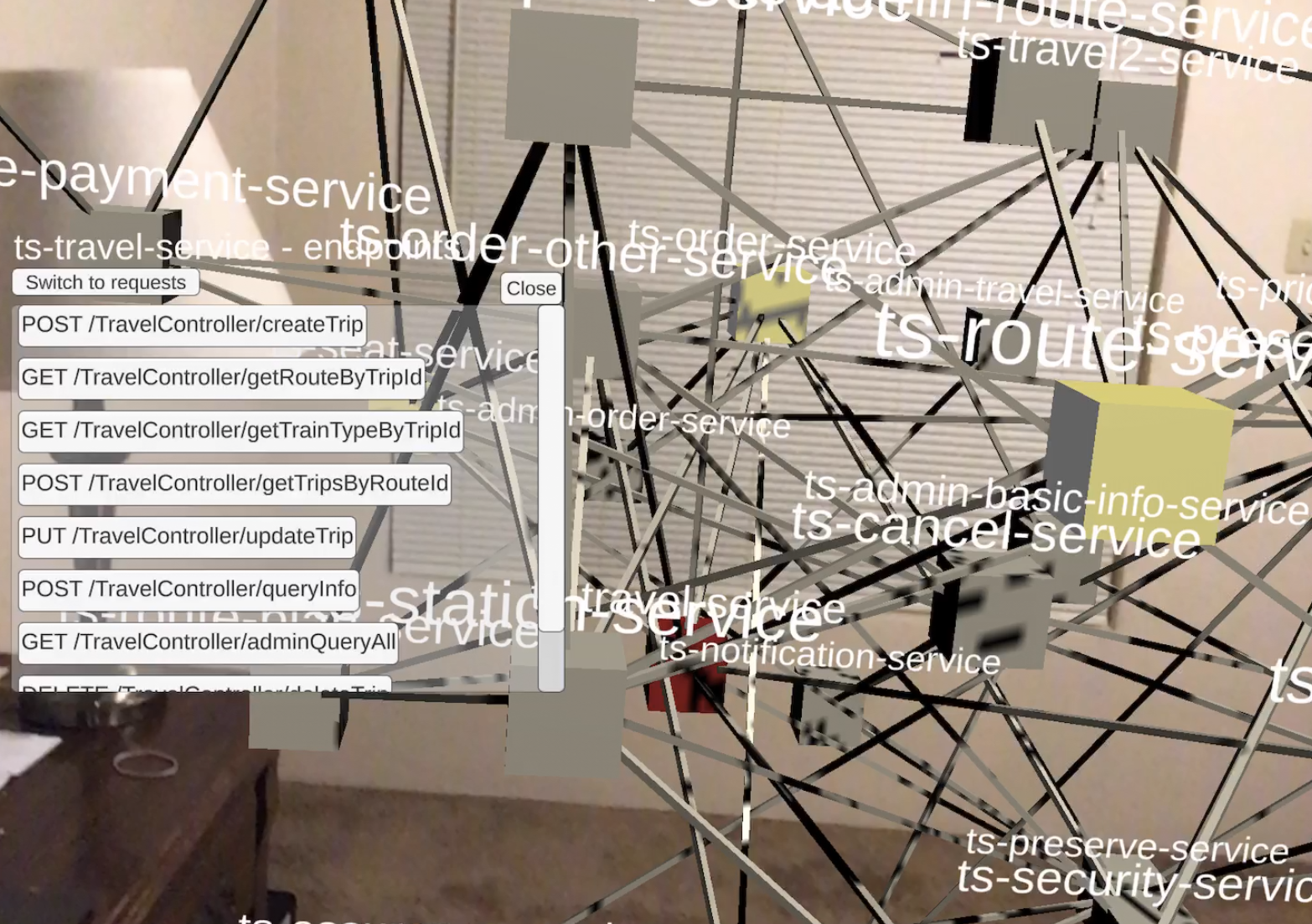}
\vspace{-.5em}
\caption{AR Tool (Complete version is available at the protocol data.)}
\label{fig:video_sample_3d}

\end{figure}

\subsection{Visualization Tools}

The study introduced two visualization tools: 2D and AR. The 2D tool is a well-established visualization similar to those used in commercial and open-source tools. It uses rectangular boxes and arrows to depict services and calls in microservice dependency graphs, as depicted in~\autoref{fig:video_sample_2d}. The AR tool, on the other hand, renders a 3D model of the service dependency graph using cubes for services and line connectors with a popup for call information, as illustrated in~\autoref{fig:video_sample_3d}. Although both visualizations provide the same information, they offer different display and interaction options that cater to different needs. The 2D approach is ideal for static and possibly printable views, while AR enables a wider interaction and visualization of multiple pieces of information in the same view.

The two visualizations have different natures in presenting the same information. The 2D tool is a web-based application that can be accessed through desktop computers and laptops. On the other hand, the AR visualization requires a device with limited screen space and a different interaction format. It is designed to be installed on mobile or tablet devices and is prepared to render the 3D representation through the mobile camera through the AR space. The AR application was distributed to Android devices through an APK file (required at least Android 8) and to iOS platforms via an IPA file (required at least iOS 11).

Both tools utilized the TrainTicket (V 0.0.1)~\cite{trainticket} testbench as the system under test, which is a microservice-based system containing 41 Java Spring microservices. The Prophet tool~\cite{bushong2022reconstructing} was used to extract the corresponding service dependency graph, and two system sizes were considered: the large system consisting of the complete 41 microservices, and the small version containing 16 microservices. The small version was generated by manipulating the graph to preserve connections between them, thereby replicating a microservices system architecture. That allows the study to examine the different system sizes' impact on tools.

\subsection{Participants}



The study population consisted of 20 participants from various institutions in the United States and Europe. Prior to the study, background information such as software engineering and microservices experience, number of microservices projects, project size, etc., was collected through a questionnaire. The questionnaire was emailed to 26 English-speaking candidates, of whom 23 were interested in participating, and eventually, 20 completed the study. Participants were selected based on having one to five years of experience in microservices-related development, which was most relevant to the study. While recruiting practitioners is challenging, the study opted to prioritize obtaining valuable outcomes for practitioners over recruiting a larger number of novices.

These participants were distributed among two groups based on the number of years of experience in microservices development. Participants who have less than two years of experience were classified as novices; otherwise, they were considered experienced. The first group (novice) consisted of nine novice developers with an average of one year of experience, while the second group (experienced) had eleven experienced developers with an average of four years of experience,~\autoref{tab:participants} shows the participants distribution. Prior to the study execution, two extra participants volunteered to participate in a pilot study, which they were not included in the actual study data. The pilot study was conducted to verify the experiment's validity. It enabled the team to modify and improve the training material, timing, tasks, and questions accordingly.

\subsection{Artifacts}
\label{sec:artifacts}

The study encompassed various artifacts, beginning with a participation questionnaire that was distributed through email to solicit candidates' interest in participating. The online form was designed to gather relevant information about the participants' experience, as well as their preferred date and time for the study. Additionally, each visualization approach included three documents that were customized to suit the corresponding content. These documents were utilized to delineate the study's execution process, as outlined below:

\begin{itemize}[leftmargin=*]
    \item \textit{Training Document}: A PDF file was used that described the corresponding tool and its required installation process in detail. It also showed an example of the anticipated tasks. A short video was included to demonstrate the tool for usage.
    \item \textit{Tasks Form}: An online form containing six questions (tasks). These tasks vary across different tools to prevent the influence of answer memorization. The tasks were of equal difficulty, and they focused on the same characteristics of various microservices, resulting in different answers. For instance, while the tasks for different versions used the same question, they were related to distinct microservices within the system. These tasks have been divided into three categories (See~\autoref{tab:tasks}): \textit{Dependency}, \textit{Cardinality}, and \textit{Bottleneck}. These categories contribute to describing and illustrating the system's understandability as mentioned in RQ.1 above.
    \item \textit{Feedback Form}: An online form asked the participants to evaluate the method after completing the tasks phase. It contained two types of questions, 5-Point Likert Scale questions as shown in~\autoref{tab:feedback-options} and paragraph questions as listed in~\autoref{tab:feedback-paragraphs} to get an in-words descriptive evaluation. These feedback guides the answers for RQ.2 and RQ.3 as stated in Section \ref{sec:rqs}.
\end{itemize}

These three documents and forms were organized into a single script document. It is a PDF file that gives a brief description of each system and the included number of microservices. Such that, it contains references to the Training Document, Tasks Form, and Feedback Form; it also gives the participants guidance about how to execute the study and how to submit the results after finished. Each method script was sent through an email to the participant in correspondence to a scheduled slot.

\begin{table}[h]
\caption{Significance Statistics for the Understandability variables}
\label{tab:rq1_stats}
\begin{tabular}{|l|c|c|c|}
\hline
\multicolumn{1}{|c|}{\textbf{Variables}} & \textbf{\begin{tabular}[c]{@{}c@{}}Dependency\\ (P-value)\end{tabular}} & \textbf{\begin{tabular}[c]{@{}c@{}}Cardinality\\ (P-value)\end{tabular}} & \textbf{\begin{tabular}[c]{@{}c@{}}Bottleneck\\ (P-value)\end{tabular}} \\ \hline
Intercept                                & \textless 0.001                                                         & \textless 0.001                                                          & \textless 0.001                                                         \\ \hline
\textbf{Experience Level}                & \cellcolor[HTML]{FFFFFF}0.986                                           & 0.371                                                                    & 0.218                                                                   \\ \hline
\textbf{System Size}                              & \cellcolor[HTML]{C0C0C0}\textbf{0.041}                                  & \cellcolor[HTML]{C0C0C0}\textbf{0.002}                                   & 0.089                                                                   \\ \hline
\textbf{Group}                           & 0.446                                                                   & 0.562                                                                    & 0.618                                                                   \\ \hline
\textbf{Session}                         & 0.067                                                                   & 0.711                                                                    & 0.640                                                                   \\ \hline
\textbf{Tool}                            & 0.515                                                                   & 0.158                                                                    & 0.556                                                                   \\ \hline
\textbf{Tool * Experience Level}         & \cellcolor[HTML]{C0C0C0}\textbf{0.035}                                  & 0.606                                                                    & 0.747                                                                   \\ \hline
\textbf{Tool * System Size}              & \cellcolor[HTML]{C0C0C0}\textbf{\textless 0.001}                        & 0.132                                                                    & 0.460                                                                   \\ \hline
\end{tabular}
\vspace{-2em}
\end{table}

\subsection{Design Procedure}

The experiment followed a 2x2 crossover design where each participant group used both visualization tools. Each participant received an email at the time slot that s/he booked for the study. This email contained two ordered scripts that the participants were required to proceed with the same order; each script contained the whole process for a specific approach as described in Section \ref{sec:artifacts}. The order of the execution was crossedover
from one participant to the other; The \textit{Group} column in~\autoref{tab:participants} indicates whether the participant starts with 2D or AR experiment. Moreover, two of the authors were on-call at each slot for supporting and answering any inquiries raised by the participants during the experiment.





The experiment duration was one hour which was decomposed as ten minutes for training, 30 minutes for executing the experiment and answering the questions (tasks), and ten minutes for filling out and submitting the feedback. Every participant was required to proceed with two sessions, each per tool. All participants received the same required knowledge and training about the experiment, the target, interfaces, study time, and methodology.

The experiment starts with the ten-minute training; which happens through the training document. The training ensures the participant was settled and familiar with the application and the nature of the tasks before proceeding to the tasks. It provided guidance for the participants to install and run the applications on their own devices and browsers. Then, it showed a use case of the usage of the application while answering a question similar to the ones in the tasks. After that, the participants proceeded with the tasks for 30 minutes to answer the tasks-related questions shown in~\autoref{tab:tasks}. The tasks' answers measure the interpretation accuracy of the system regarding the three perspectives of dependency, cardinality, and bottleneck. Therefore, a percentage of correctness was evaluated for each participant referencing the actual answers extracted from the testbench system. This data is analyzed to show the impact of the study variables on the understandability of the system.



Finally, once the tasks are submitted, the feedback step is highlighted. It consists of two parts: usage-related criteria and open-opinion questions. The first part questions are shown in~\autoref{tab:feedback-options}, they use an ordinal 5-point Likert scale (1 is the worst, 5 is the best). These data are statistically analyzed in the upcoming sections. The second part requires the participants to write down their opinions for answering the question in~\autoref{tab:feedback-paragraphs}. This feedback was analyzed to extract common thoughts and recommendations.


The study aimed to create balanced groups based on system size and participant experience, with ten small and ten large systems, and a specific number of novice and experienced participants assigned to each system. To prevent one tool from exerting more influence than the other, we used abstract names (such as Tool 1 and Tool 2) for the two tools, to minimize any impact that names may have on participants. Additionally, we randomly ordered the two scripts for the two tools so that some participants in each group completed the study in reverse order from others. The order of tool execution for each participant is indicated in the \textit{Session 1} and \textit{Session 2} columns of \autoref{tab:participants}, where \textit{Group} column denotes participants who started with 2D as G1 and those who started with AR as G2.

Due to compatibility issues with six participants' mobile devices, they struggled to use the AR tool. It resulted in a lack of AR tool results for four small and two large participants. Therefore, the participant assignments were slightly adjusted, as depicted in \autoref{tab:participants}. Specifically, two novice participants were moved from the large to the small system, and one experienced participant was moved from the small to the large system, as highlighted in bold in the Participant column. Additionally, 14 participants completed both sessions, such that G1 conducted five large and three small experiments, while G2 carried out three large and three small experiments.

To sum up, the assessment of the procedure indicated that all participants were able to carry out the 2D experiment successfully. Nonetheless, six participants could not install the AR tool because their mobile devices did not meet the OS requirements (minimum Android 8 or iOS 11). Despite this, their results were still included in the 2D tool outcomes, but they were considered missing data for the AR tool. All of the received email inquiries from participants were related to the AR tool's installation issue, and otherwise, the study was clear to proceed without any further concerns.

\subsection{Analysis Approach}


First, we will report descriptive statistics to continue with hypothesis testing. Different tests are used for RQ1 and RQ2. 

\subsubsection{RQ1}

Following Vegas et al.~\cite{7192651}, we use the Linear Mixed Effects Model to conduct the analysis for RQ1 in this experiment. 
It is used to test the hypotheses of the effect of the tool (2D/AR), the experience level of the participant (novice/expert), and system size (small/large), along with the two-way interactions of experience level by tool and system size by the tool on our dependent variables: dependency, cardinality, and bottleneck, controlling for possible effects derived from the chosen crossover design (session~and~group).

We used SPSS v.27 MIXED procedure to analyze our data. Our data were fitted using two representations of the data: one that takes into consideration participant as a random factor, and another one that takes into consideration that the tool is a within-subjects factor. For the latter case, five different covariance structures (identity, diagonal, first-order autoregressive, compound symmetry, and unstructured) are fitted. This makes six different models to be fitted for each independent variable.

We used Akaike’s Information Criterion (AIC) to measure model fit. AIC values are not large or small per se, but their values can be compared across models. Smaller values mean better fitting, so, we will choose the model with the lowest AIC.

The MIXED procedure requires the normality of residuals. We checked this by means of normal probability (Q-Q) plots. When data do not meet this criterion, transformations of the dependent variable need to be used.

The effect size is also reported in terms of Cohen’s d statistic~\cite{cohen2016power} if we can ensure that the data follows a normal distribution; otherwise, we report non-parametric effect size, Cliff’s delta~\cite{cliff1993dominance}.

\subsubsection{RQ2}
We use the non-parametric Wilcoxon signed rank test for related samples to analyze RQ2. It is used to test the hypotheses of the effect of the tool (2D/AR) on our dependent variables: easy to use, easy to understand, time to get the information, and has all information needed.

\subsubsection{RQ3}

We performed a qualitative analysis for the short answers that were submitted regarding the questions in~\autoref{tab:feedback-paragraphs}. We read all the received feedback on the 2D and AR experiments. For objectivity, each of us categorized the feedback based on the interpretation of each answer individually. Then, they discussed and combined similar categories. After that, they summarized and reported the common feedback.

\section{Data Analysis Results}

\begin{figure*}[!ht]
\vspace{0pt}
\begin{minipage}[b]{0.45\linewidth}
\centering
\includegraphics[width=\textwidth]{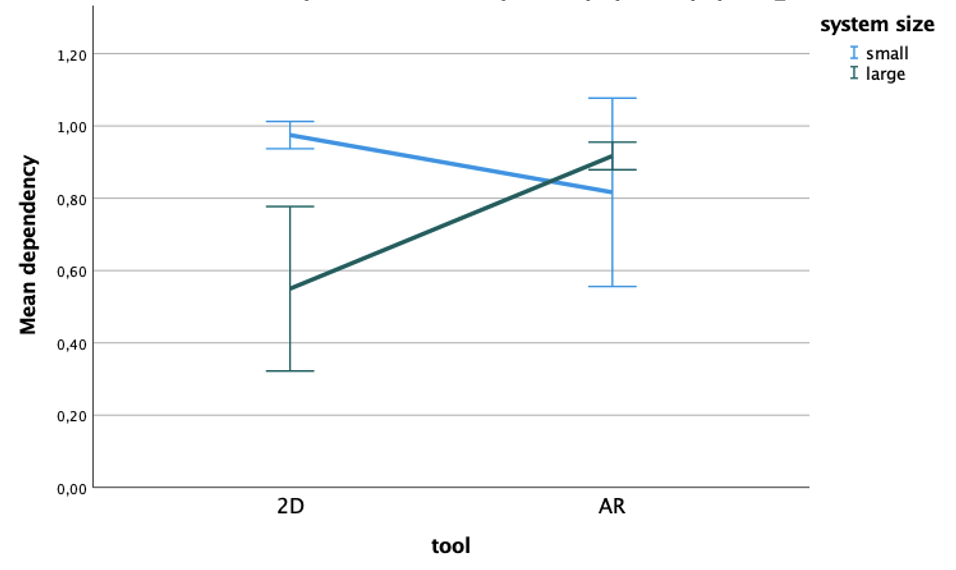}
\vspace{-2em}
\caption{The impact of system size on the dependency ($\delta$ = 0.96)}
\label{fig:systemsize_tool_dependency}
\end{minipage}
\hspace{0.5cm}
\begin{minipage}[b]{0.45\linewidth}
\centering
\includegraphics[width=\textwidth]{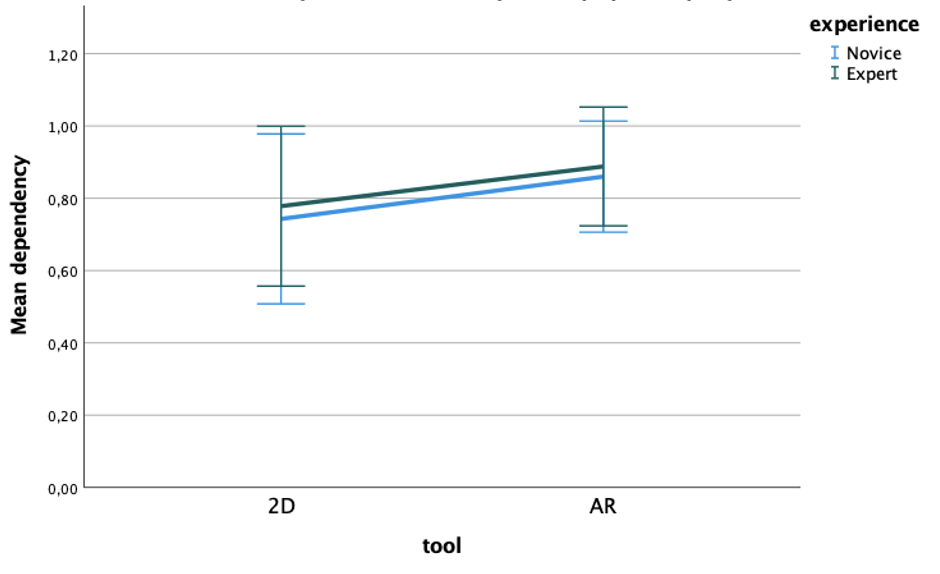}
\vspace{-2em}
\caption{The impact of experience level on the dependency ($\delta$ = 0.19)}
\label{fig:experience_tool_dependency}
\end{minipage}
\vspace{-1.5em}
\end{figure*}





\begin{figure}[ht]
\centering
\includegraphics[width=0.67\textwidth]{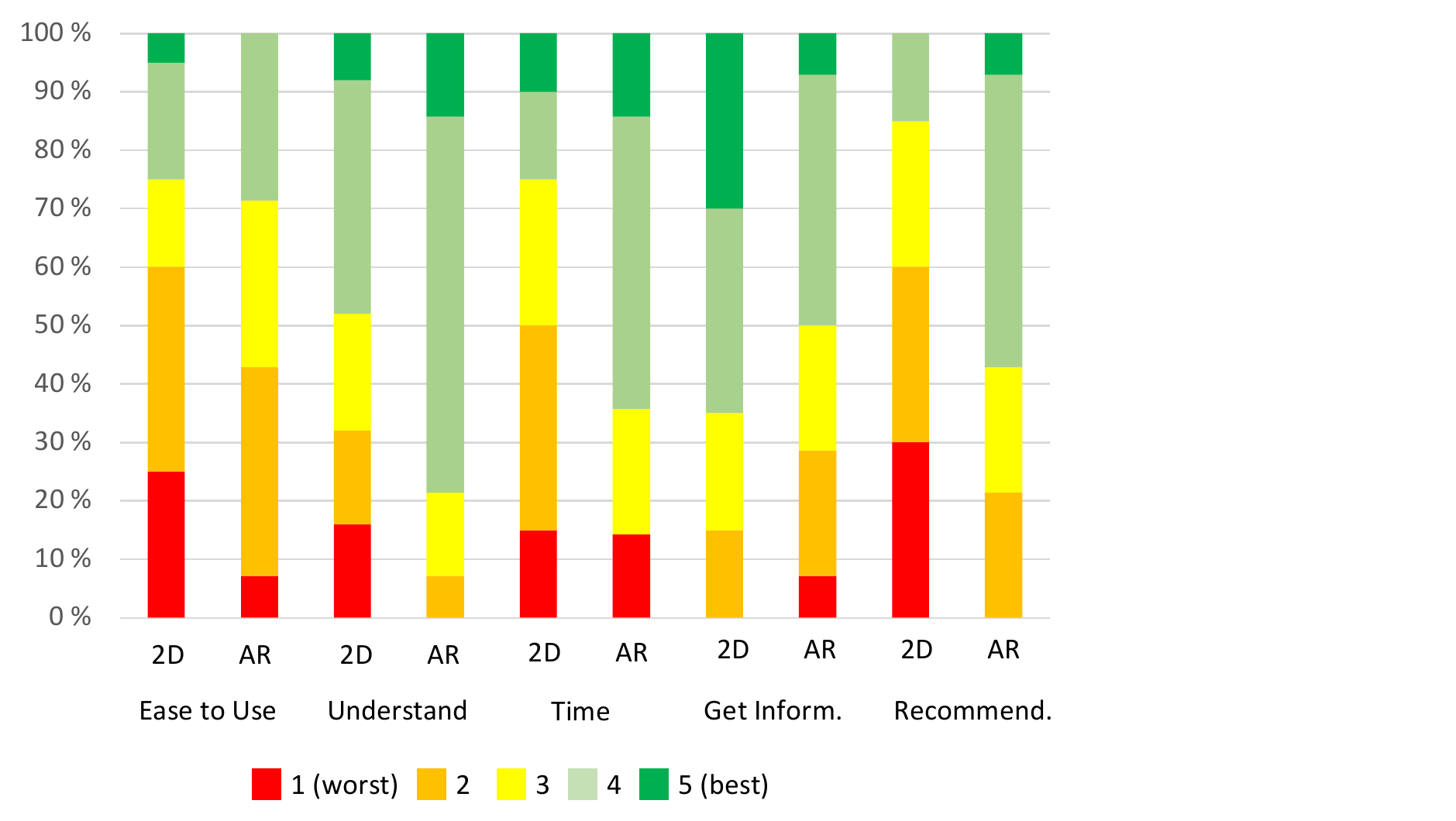}
\vspace{-2em}
\caption{What is the verbalized perception regarding AR and 2D (RQ2)}
\label{fig:stackedRQ2}
\vspace{-1.5em}
\end{figure}



We analyzed the dataset\footnoteref{foot-protocol} collected from 20 participants. Executing the analysis approach per each corresponding data to answer the RQs. 

\subsection{RQ1. Is 2D visualization more applicable than AR for
understanding Microservice-based systems?}
We examined six models to measure how they fit the data. These models are configured based on how participants are modeled, such that one model has no repeated measures specified, then it specifies the participants as a random variable. On the other hand, the other five models specified repeated measures. However, they examined different covariance types, such as scaled identity, compound symmetry, diagonal, unstructured, and autoregressive.

We compare the fitting degree of these models based on AIC and normality of residuals (observing the Q-Q plots). As a result of this comparison, the model configured by the participants as a random variable is chosen; It shows the normality of residuals (thus, data transformation is not needed) and it has the best AIC value among all the others. This model is applied to dependent variables that represent the understandability property of the study as follows: dependency, cardinality, and bottleneck. 

Examining the significance of independent variables over the understandability is summarized through the p-value as listed in~\autoref{tab:rq1_stats}. The dependent variable is influenced by the interaction between the tool and each system size and experience level. Although the system size variable shows significance, we do not consider its effect because its interaction with the tool has more precedence. To measure this impact, we plot the estimated marginal means for the interaction of the variables and calculate their effect size. First, the interaction between system size with the tool emphasizes the scalability of the AR tool. The AR tool appeared as the best choice for visualizing large systems with large effect size ($\delta$ = 0.96), however, there is no large impact on the small systems from both tools as illustrated in~\autoref{fig:systemsize_tool_dependency}. Following the same for the experience level significance, the analysis shows that the 2D tool requires more experienced participants in order to detect the dependency of a system. Nevertheless, it is a small effect size ($\delta$ = 0.19) as shown in~\autoref{fig:experience_tool_dependency}.

The cardinality variable highlights the system size as its only impact factor, while the AR tool does not have an impact on the accuracy of the cardinality-related questions. However, the analysis shows that the small enables the participants to answer the cardinality questions more accurately. Furthermore, the AR does not have an impact on the bottleneck results, thus, either the 2D or the AR tool can be chosen for these related tasks without influencing the results.







\subsection{RQ2. What is the verbalized perception of the participants regarding the use of AR for understanding microservice-based systems?}

The easy-to-use scores were compared for both tools. On average, the 2D tool performed worse (Mdn = 2) than the AR tool (Mdn = 3), with a small effect ($\delta = 0.19$). The Wilcoxon signed-rank test indicated that this difference was not statistically significant (T=30, Z=0.263, p = 0.793).

The easy-to-understand scores were compared for both tools. On average, the 2D tool performed worse (Mdn = 3) than the AR tool (Mdn = 4), it showed a large effect ($\delta = 0.46$), especially with the large system. The Wilcoxon signed-rank test indicated that this difference was statistically significant (T=66, Z=2.164, p = 0.030). 

The time-to-get-information scores were compared for both tools. On average, the 2D tool performed worse (Mdn = 2.5) than the AR tool (Mdn = 4), with a large effect ($\delta = 0.38$). The Wilcoxon signed-rank test indicated that this difference was not statistically significant (T=67, Z=1.548, p = 0.122).

The has-all-information-needed scores were compared for both tools. On average, the 2D tool performed better (Mdn = 4) than the AR tool (Mdn = 3.5), with a small effect ($\delta = 0.28$). The Wilcoxon signed-rank test indicated that this difference was not statistically significant (T=16, Z=-1.218, p = 0.223).

Recommendation scores were compared for both tools. On average, the 2D tool performed worse (Mdn = 2) than the AR tool (Mdn = 4), with a large effect ($\delta = 0.57$). The Wilcoxon signed-rank test indicated that this difference was statistically significant (T=69, Z=2.435, p = 0.015).

Moreover, we applied the descriptive statistics to these response variables as summarized in~\autoref{fig:stackedRQ2}. Both AR and 2D tools do not show a significant impact on the easy-to-use response (\autoref{fig:stackedRQ2} Ease to Use). The 2D shows better ranks than AR for the has-all-information-needed variable as shown in~\autoref{fig:stackedRQ2} (Get Inform.). However, the AR tool outperforms the 2D tool for both the easy-to-understand (\autoref{fig:stackedRQ2} Understand) and the less time-to-get-information (\autoref{fig:stackedRQ2} Time). In addition, participants highly recommend the AR over the 2D as illustrated in~\autoref{fig:stackedRQ2} (Recommend.).


\subsection{RQ3. What are the challenges perceived by the participants with regard to AR-based techniques?}

Analyzing the participants' evaluation and answers for the questions listed in~\autoref{tab:feedback-paragraphs}; their responses are summarized in~\autoref{tab:feedback_comparison}. For the 2D tool, the participants highlighted the pain of scrolling a lot around the graph to follow the crossing lines between services. They recommended showing the information as needed, such as when the user clicks on a specific node or edge. 
For the AR tool, participants indicated that they needed more control to be able to zoom in and rearrange the service nodes. In addition, improved usage of colors was a common recommendation for both approaches, either to distinguish between different request types or between the dependency directions. The participants indicated that they were highly impacted by the complexity of tracking the dependency lines in the 2D tool, in contrast to the simpler dependency lines in the AR tool. They also stated that the AR improves their understanding of the system by showing an overview picture of the system and clear dependencies between services.

\begin{table*}[h]
\centering
\vspace{0em}
\caption{Paragraph Questions Feedback Analysis (RQ3)}
\label{tab:feedback_comparison}
\vspace{-0.5em}
\begin{tabular}{p{30em}p{35em}}
 \textbf{2D Tool}  & \textbf{AR Tool} \\
 \hline
\multicolumn{2}{|c|}{\textbf{What has been your greatest pain point?}} \\
 \hline
   \makecell*[{{p{30em}}}]{1. Scrolling around the diagram.\\
2. Many crossing communication lines.\\\,} & \makecell*[{{p{30em}}}]{1. Move a lot around the graph.\\
2. AR Engine shifts the graph suddenly.\\
3. The fixed location for the services.}\\
 \hline

\multicolumn{2}{|c|}{\textbf{What is your recommendations for enhancing this method?}}  \\  
 \hline
 \makecell*[{{p{30em}}}]{1. Highlighting the dependency using colors.\\
2. Show information as needed.\\
3. Provide Search and filter features.\\
4. Dialog box to show the dependent and depending services. }     &  
\makecell*[{{p{35em}}}]{1. Provide control to zoom and relocate the nodes.\\
2. Use a different color for the dependency on the other direction.\\
3. Allow color customization.}
 \\
  \hline
\multicolumn{2}{|c|}{\textbf{What is the information that you can easily get using this method?}} \\
 \hline
 \makecell*[{{p{30em}}}]{1. Endpoint Information.\\
2. The services dependencies cardinality.}  &  
\makecell*[{{p{30em}}}]{1. System overview picture.\\
2. Supported Endpoints per service.\\
3. Dependency links between services.}
\\
 \hline
\multicolumn{2}{|c|}{\textbf{What is the information that you find hard to get from this method?}}\\
 \hline
\makecell*[{{p{30em}}}]{1. The service dependencies.\\
2. The direction of the dependency.\\\,} & \makecell*[{{p{30em}}}]{1. Invoked Endpoint details.\\
2. The services that depend on specific service \\(The other dependency direction).}       \\                                                                            
\end{tabular}
\vspace{-2.8em}
\end{table*}


\section{Discussion}
The results of this work enabled us to understand the power of Augmented Reality for understanding microservice architecture. 


When evaluating the suitability of AR for detecting service dependencies, experienced developers performed similarly using both tools. However, AR enables novice developers to achieve a level of understanding comparable to experienced developers when dealing with small and large systems. As a result, identifying system dependencies using 2D requires a greater level of expertise compared to AR. One of the reasons for the improved performance of novice developers, especially in the large system, is that these systems are usually harder to understand, and AR tools enable them to rotate them three-dimensionally to zoom or better focus on a specific portion of the system, while 2D systems might be complex for novices possibly because of the condensed amount of information rendered into limited space. Furthermore, the analysis shows that the system size has a significant impact on detecting the dependencies using the studied tools. Such that, the 2D tool could visualize this aspect clearer than the AR tool for small systems. However, AR outperforms visualizing large systems in a scalable behavior.

Unexpectedly, when analyzing the cardinality of services, no differences emerged between 2D and AR. Participants, independently from their expertise, were able to identify cardinality accurately in both visualizations. However, as expected, the cardinality resulted in being associated with system size. In a larger system, the identification of cardinality was more complex and required more time.

The identification of the bottleneck is also, unexpectedly, not influenced by the visualization approach adopted, nor from the system size or experience of the participants. 

When considering the opinion of the participants on the 2D and AR visualizations (RQ2), the AR tool enables the developers to better understand the microservice-based system analyzed. The reason could be because of the 3D visualization of the AR tool that supports different viewpoints in higher rendering space. Furthermore, the recommendation results confirm the preference for the AR tool due to the clearer dependency representation adopted in the AR tool.

The ease to use, the time needed to get information, and the information needed by the participants to analyze the system  do not provide statistically significant results. However, even if the results are not significant, the time needed to get information to AR was way lower than the 2D tool in most cases, and the 2D tool provided more information than AR for analyzing microservice-based systems. That could be because the AR visualization requires extra steps to find the needed information.

In summary, we recommend novice developers consider AR visualization to visualize dependencies in large systems. However, 2D visualization of the service call graph (e.g., the visualization provided by Jaeger or Kiali) might still be useful to identify service cardinality and system bottlenecks. 

Therefore, considering the lack of an experienced workforce in the software development of microservice systems, we highly recommend companies and researchers to further develop and validate the AR approaches for visualizing microservice-based systems. That also promotes the onboarding process of new joiner developers in microservice-based projects. On the other hand, while the participants recommended the AR tool more than the counterpart (\fig{stackedRQ2} - Recommend.), it must be considered that a hello effect (first impression) could occur upon seeing an unconventional approach. It is important to underline that for AR to become successful, multiple features and common expectations from the feedback will need to be incorporated into the method. For instance, users demand a quick mechanism to search and filter. Although AR seems promising, it adds restrictions for practical usages, such as requiring developers to use portable devices with cameras. Also, that might be somewhat uncomfortable being apart from their work environment tools. Therefore, considering the integration of the visualization tool with the development workspace could enhance the usability of these tools.

\section{Threats to Validity}



We address the threats to the validity of our research according to the classification proposed by 
\cite{wohlinThreats}.

\subsubsection{Construct Validity}


This study followed the control experiment practices and guidelines to construct, analyze, and report the study and its results. The core of the study involved our prototype tools, which could affect the results. However, both prototype tools were developed with the best intentions toward usability. The prototype tool authors were distinct from those who designed the user study to avoid bias. There was no communication about the study design and execution between them. Furthermore, these tools rendered the results produced using our SAR process; nevertheless, this SAR-related software is published and validated through multiple projects.

We collected results using small but realistic tasks expected from practitioners, which are chosen based on interviews with experienced developers in microservices.

\subsubsection{Internal Validity}




Potential threats to the internal validity of the experiment are fatigue, carry-over, and practice effects. The fatigue effect is considered negligible, as the task execution was relatively short. The study was performed remotely with two authors available on call/email at allocated experiment times; the participants chose their most convenient time to proceed with the study. Carry-over and practice effects were desirable and unavoidable in this experiment due to its nature, i.e., subjects adopted 2D and AR consecutively, and they may have transferred their learning of the identification of dependencies, cardinality, and bottlenecks. We adopted a 2-group crossover within-subjects design to avoid the carryover effect. We used different questions involving another part of the system accomplishing the same goal to limit the memorization of answers. To avoid user bias, we referenced both studies with agnostic names, so the participant would not predict or associate with another applied method.

Furthermore, the two visualization tools present distinct display and interaction options due to their inherent characteristics. While the 2D tool lacks interactivity, the AR tool's use of pop-ups could potentially have an effect. Nonetheless, in the participants' feedback, no significant difference was observed in terms of the ease of solving their tasks. This aligns with the study's objective of assessing whether the AR approach can outperform the traditional 2D approach in typical microservice evaluation tasks despite its different nature and the devices' small size limitations.





The study process faced compatibility threats that prevented some participants from installing the AR tool resulting in an unbalanced number of study execution for 2D than AR experiments. Moreover, the participants executed the study in the reverse order that it was sent to them, and they impacted the distribution of the crossover across the study. However, the statistical results proved that performing the tasks with one visualization followed by a similar set of tasks for the second visualization should not have an impact on the study results. The participants' mobile devices hold specific threats to the AR-related study since different screen sizes of their devices could impact their understandability of the system. However, there was a lack of distribution and not enough categorized data for screen size to include and examine its impact~as~a~factor~in~the~model.

The threats related to participants' understanding of the process and the tasks were considered in the feedback form. Such that most of them stated that was clear; however, a few participants mentioned that the tasks were not clear enough from their perspective. That could also be because of a lack of their experience. Therefore, we addressed the experience level as one of the main factors of the study.

Finally, The study was limited by the number of participants, we chose to prioritize meaningful results for practitioners over a larger group of students. The study followed a within-subjects design, allowing each participant to work with both tools and effectively increasing the number of participants from 20 to 40. Moreover, the lack of power in the number of participants reduced the significance of some measurements, including bottleneck detection measurements. Nevertheless, that requires further investigation using more participants to ensure its significance.




\subsubsection{External Validity}


We selected participants among our contacts by directly contacting them. However, we reached a diverse spectrum of practitioners from North America and Europe, thus from various backgrounds. We surveyed study participants about their previous experience and relied on the trustworthiness of their answers. In the end, the subject sample was quite heterogeneous, from junior to senior levels concerning their experience of microservices. Moreover, we have assigned the study in a simple random method to avoid biased results.

To avoid participant familiarity with the studied tools and system we considered presenting it to them for the first time. We have used a third-party benchmark recognized by the community to avoid bias. We manually manipulated the system to produce the small and large perspectives of the system size. However, we randomized the distribution of~both~variants~among~participants.

\subsubsection{Conclusion Validity}

To avoid bias, we included various authors from different fields and expertise. The results of the study collected the opinions using natural texts and Likert scale answers, then these data were statistically and descriptively analyzed. Moreover, the primary goal of this research was to answer three research questions, and the user study generated additional insights on practical implications, limitations, and open challenges which need to be addressed when transitioning the AR tool to practice.

\section{Conclusion and Future Work}






We conducted a controlled experiment to compare the perceived system understandability of microservice-based systems from a service-call graph represented in 2D and Augmented Reality (AR). The study included multiple realistic tasks to measure the ability of the developers to understand the system using the studied tools. It highlighted the impact of the participants' expertise and the different system sizes. The tasks' answers and the feedback were statistically and descriptively analyzed.

The findings investigate the feedback regarding the different visualization aspects and information that should be provided to better systems understandability and to further develop the AR system.
We conclude that AR visualization enables novice developers to understand system dependencies as if they were experienced developers. However, experienced developers do not get significant benefits from AR for analyzing some of the tasks. One of the reasons for the latter could be that experienced developers are used to assessing systems using 2D visualizations, and the introduction of AR might have biased their performance. On the other hand, AR emphasizes a large impact on scaling and visualizing large systems; however, it could not be significant for small systems visualization.
%


In future work, we will replicate this study with senior developers only and analyze industrial systems to better understand the usefulness of AR but also the information needed by developers to understand the system. We will consider interactive 3D visualization as an alternative as well. Dynamic system perspectives can also benefit from advanced visual perspectives.

\section*{Acknowledgements}
This material is based upon work supported by the National Science Foundation under Grant No. 1854049 and a grant from \href{https://research.redhat.com}{Red Hat Research}, a grant from the Ulla Tuominen Foundation (Finland), a grant from the Academy of Finland (grant n. 349488 - MuFAno), and project PGC2018-097265-B-I00, funded by: FEDER/Spanish Ministry of Science and Innovation---Research State Agency.

\bibliographystyle{IEEEtran}
\bibliography{access}

\end{document}